\documentclass[PRB,twocolumn,showpacs,amsmath,superscriptaddress]{revtex4-1}
\usepackage[colorlinks,urlcolor=blue,linkcolor=blue,citecolor=blue]{hyperref}
\usepackage{overpic}
\usepackage{subfigure}
\usepackage{dcolumn}
\usepackage{latexsym}
\usepackage{amsfonts}
\usepackage{color}
\usepackage{bm}
\usepackage{array}
\usepackage{booktabs}
\usepackage{caption}
\usepackage{multirow}
\newcommand{\tabincell}[2]{\begin{tabular}{@{}#1@{}}#2\end{tabular}}
\makeatletter

\newcommand{\Rmnum}[1]{\expandafter\@slowromancap\romannumeral #1@}
\makeatother

\begin{document}

\title{Topological pairings in Janus monolayer TaSSe}

\author{W. P. Chen} 
\affiliation{
	National Laboratory of Solid State Microstructures And Department of Physics, Nanjing University, Nanjing 210093, China}

\begin{abstract}
	The Janus monolayer transition metal dichalcogenides[TMDs] MXY[M=Mo,W, etc. and X,Y=S,Se, etc.] has been synthesized recently, and the Rashba spin splitting arises in it owing to the breaking of out-of-plane mirror symmetry[\href{https://journals.aps.org/prb/abstract/10.1103/PhysRevB.97.235404}{Phys. Rev. B 97, 235404 (2018)}]. Here we study the pairing symmetry of superconducting Janus monolayer H-TaSSe by solving the linearized gap equation at the critical temperature $T_c$. We find that the strong Rashba effect in H-TaSSe could produce topological superconducting states which differs from that in its parent monolayer H-TaS$_2$ and H-TaSe$_2$. More specifically, at the chemical potential $\mu>-0.08$eV, we obtain a time-reversal invariant s+f+p-wave mixed pairing state. This pairing state is full-gap and topologically nontrivial, i.e. $\mathbb{Z}_2=1$. However, a time-reversal broken d+p+f pairing belonging to the 2-dimensional irreducible representation $E$ appears at lower chemical potential. It can host a large Chern number $C=-6$ at appropriate pairing strengths. The results suggest the monolayer H-TaSSe to be a candidate helical or chiral topological superconductor. 
\end{abstract}

\maketitle
\section{Introduction}
 Topological superconductors(TSCs), which are characterized by a full-gap bulk and robust gapless surface states, have gained much attention for their novel topological properties recently\cite{Fu2008,Schnyder2008,Kitaev2009,Fu2010,Ryu2010,Qi2011,Das2012,Fidkowski2013,Sato2017}. Several superconductors are considered to be unconventional and even topologically nontrivial, including Sr$_2$RuO$_4$\cite{Rice1995,Baskaran1996,Nelson2004} and some heavy fermion compounds\cite{Stewart1984,Varma1985,De1987}. There are also theoretical proposals for realizing the topological superconducting states artificially by using the s-wave superconductor and spin-orbital coupling(SOC). For instance, through proximity effect, the s-wave superconductor on a strong topological insulator becomes a spinless p+ip-wave superconductor\cite{Fu2008}. Besides, by applying a strong perpendicular magnetic field to an s-wave Rashba superconductor, topological superconducting states can emerge\cite{Sato2009,Sau2010}.  However, there's no definitive experimental evidence for TSCs until now, so it's still a challenge to search for them theoretically and experimentally.

 Transition metal dichalcogenides(TMDs) are one kind of noncentrosymmetric layered van der Waals materials, which have been studied for decades as they always exhibit superconductivity, CDW and other electronic phenomena\cite{Clayman1971,Boaknin2003,Huang2007,Berthier1976,Mutka1983,Neto2001}. Superconductivity in the bulk TMDs is generally thought to be conventional BCS-type. However, in the ultra-thin films, a strong enhancement of the in-plane critical field indicates possible unconventional superconductivity caused by the large Ising SOC \cite{Lu2015,Xi2015,Saito2016,Xing2017,Lu2018,Barrera2018}. Two kinds of parity-mixed pairings, i.e. s+f-wave and d+p-wave, have been suggested as candidate unconventional superconducting states in the monolayer TMDs\cite{Yuan2014,Hsu2017}. 

 This work is on the basis of newly synthesized Janus TMDs MXY(M=Mo,W,Ta, etc. and X,Y=S,Se,Te, etc.)\cite{Lu2017,Zhang2017,Li2018,Hu2018,Shi2018,He2018,Zhou2019,Yag2019}. In a Janus TMD molecular layer, M atomic plane is sandwiched by two different atomic planes X and Y, which breaks the out-of-plane mirror symmetry, leading to a Rashba-type spin splitting in the band around $\Gamma$. Previous works on the Janus layered materials mainly focused on the electronic structure, while the superconducting pairing phase of them has not been discussed. We choose H-TaSSe as our present pairing-symmetry study object based on the following facts: First, its parent phases H-TaS$_2$ and H-TaSe$_2$ are intrinsic superconductors. Thus, H-TaSSe is very likely to be a superconductor too. Second, there are $\Gamma$-centered FS sheets in the normal states of H-TaS$_2$ and H-TaSe$_2$\cite{Rossnagel2005,Sanders2016}. Similar FS sheets in H-TaSSe then will be subjected to a large Rashba spin splitting, which could affect the pairing symmetry. Therefore, because of the strong SOC effects, including both the Rashba and Ising type, singlet and triplet pairings would be significantly mixed together, and topological superconducting states differing from that in monolayer H-TaX$_2$ may be achieved in Janus monolayer TaSSe. We simply ignore CDW here for the suppression of it in the 2D TMDs\cite{Navarro2016,Yang2018}. 

 By solving the linearized gap equation at the critical temperature, we obtain the most favorable mixed pairing phases of Janus monolayer TaSSe at different pairing interactions and chemical potentials. Our results demonstrate that, at relatively high chemical potential, a time-reversal invariant(TRI) s+f+p-wave pairing state of irreducible representation(IR) $A_1$ will dominate the nearest-neighbor(NN) pairing channels. Compared to the nodal s+f-wave pairing of monolayer H-TaS$_2$, this s+f+p-wave pairing has a large additional $p\pm ip$-wave component induced by the Rashba SOC, making the superconductor to be full-gap and topologically nontrivial. At a lower $\mu$, however, a time-reversal broken(TRB) d+p+f-wave pairing belonging to 2-dimensional(2D) IR $E$ appears in the phase diagram. It can be a large-Chern-number[$C=-6$] chiral TSC at appropriate pairing strengths. On the other hand, the TRB d+p-wave pairing of monolayer H-TaSe$_2$ always hold a trivial Chern number. This difference could be owing to the Rashba-induced non-unitary p- and f-wave pairing components in the $E$ phase of H-TaSSe.

\section{Model} 
 
 \begin{figure}[t]
 	\includegraphics[width=8.5cm]{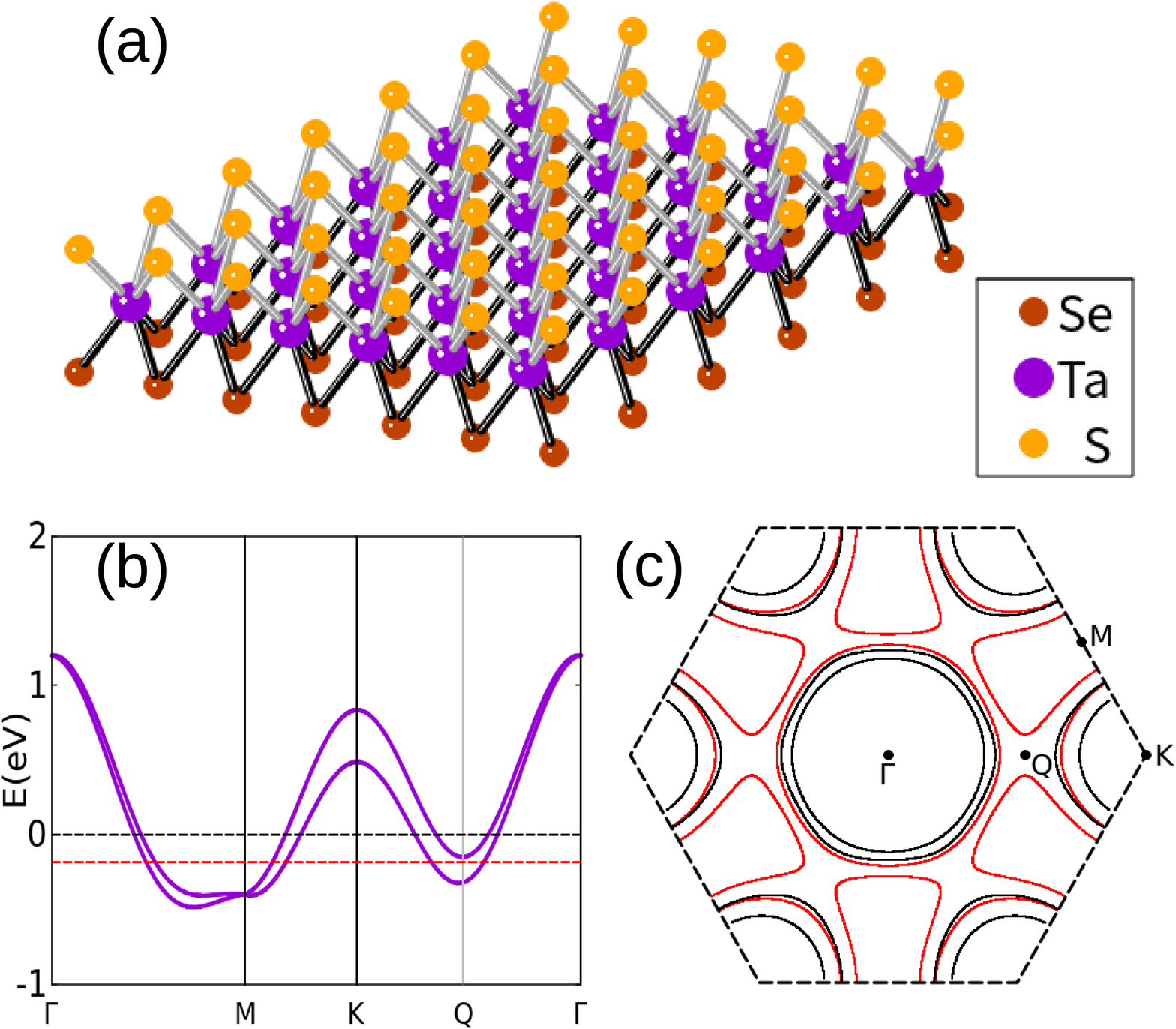}		
 	\vspace{-0.0cm}
 	\caption{\label{fig01}.
 		(a) Crystal structure of monolayer H-TaSSe, (b) the constructed band structure of it. Fermi levels at $\mu=0$ eV and $-0.18$ eV are denoted by black and red dashed lines, respectively, and their corresponding FS are presented in (c).
 	} 
 \end{figure}

The crystal structure of monolayer H-TaSSe is shown in Fig. \ref{fig01}(a). It can be viewed as the monolayer H-TaS$_2$[H-TaSe$_2$] whose bottom[top]-layer S[Se] atoms are replaced by Se[S]. Given the absence of literatures about H-TaSSe, and note that it will preserve robust electronic structure from its parent phases\cite{Hu2018}, we construct its normal Hamiltonian based on H-TaS$_2$ and H-TaSe$_2$. Thus a tight-binding model describing both H-TaS$_2$ and H-TaSe$_2$ can be given by
\begin{eqnarray}
&h_0(\bm{k})&=\xi_{\bm{k}}+\bm{\beta}_{\bm{k}} \cdot \bm{\sigma}, \\ 
&\xi_{\bm{k}}&=-\sum\limits_{j=1}^{3} [2t \cos k_{j}
+2t_1 \cos(k_{j}-k_{j+1})] -\mu, \\
&\bm{\beta_k}&=\beta_{so}\bm{z}(\sin k_1+ \sin k_2+ \sin k_3)
,\end{eqnarray}
where $k_j=\bm{k}\cdot \bm{R_j}$, with bonding vectors $\bm{R}_1=a\bm{x}$, $\bm{R}_2=a(-\frac{1}{2}\bm{x}+\frac{\sqrt{3}}{2}\bm{y})$, $\bm{R}_3=a(-\frac{1}{2}\bm{x}-\frac{\sqrt{3}}{2}\bm{y})$ and $\bm{R}_4\equiv\bm{R}_1$, $a$ the lattice constant. $t, t_1$ denote the NN and next nearest-neighbor(NNN) hopping. $\mu$ is the chemical potential and $\beta_{so}$ the Ising SOC parameter. We fix $(t, t_1, \beta_{so})=(-0.06,-0.14, 0.067)$eV, and set the chemical potential $\mu$ as an adjustable parameter. When $\mu=0$eV, this model can fit the band structure of H-TaS$_2$ given by experiments and DFT calculations very well\cite{Navarro2016,Sanders2016,Zhao2017}. On the other hand,  a chemical potential $\mu=-0.18$ eV will give a FS topology of H-TaSe$_2$, which has dog-bone-shaped electron Fermi pockets centered at $M$\cite{Rossnagel2005}. 
  
  \renewcommand\arraystretch{1.5}
  \begin{table}[t]
  	\centering
  	\caption{Classification of basis gap functions based on $C_{3v}$ symmetry. They are introduced as: $C(\bm{k})=\frac{1}{\sqrt{3}}\sum \limits_{i=1}^3\cos{k_i}, C_+ (\bm{k})=\frac{1}{\sqrt{3}}\sum \limits_{i=1}^{3}\omega^{i-1} \cos{k_i}, S(\bm{k})=\frac{1}{\sqrt{3}}\sum \limits_{i=1}^{3}\sin{k_i}, S_+ (\bm{k})=\frac{1}{\sqrt{3}}\sum \limits_{i=1}^{3}\omega^{i-1}\sin{k_i}, C_- (\bm{k})=C_+^* (\bm{k}), S_- (\bm{k})=S_+^* (\bm{k})$, with $\bm{x}_\pm=(\bm{x}\pm i\bm{y})/2$ and the phase factor $\omega=e^{-i\frac{2\pi}{3}}$.}
  	\begin{tabular}{cccc}
  		\hline
  		\specialrule{0.05em}{3pt}{3pt}
  		$\Gamma$ && Singlet & Triplet  \\
  		\specialrule{0.05em}{3pt}{3pt}
  		$A_1$ && \tabincell{c}{$\Psi^{A_1,on}_{}=\frac{1}{\sqrt{2}}$ \\  $\Psi^{A_1,nn}_{}=C(\bm{k})$} & \tabincell{c}{$\bm{d}^{A_1,z}_{}=S(\bm{k})\bm{z}$ \\$\bm{d}^{A_1,xy}_{}=i[S_-(\bm{k})\bm{x}_+-S_+(\bm{k})\bm{x}_-$]}\\
  		\specialrule{0.05em}{3pt}{3pt}
  		$A_2$ && & $\bm{d}^{A_2,xy}_{}=S_-(\bm{k})\bm{x}_++S_+(\bm{k})\bm{x}_-$\\
  		\specialrule{0.05em}{3pt}{3pt}
  		E & &  \tabincell{c}{$\begin{cases} \Psi^{E,nn}_{1}=C_+(\bm{k})    \\ \Psi^{E,nn}_{2}=C_-(\bm{k})\end{cases}$} & \tabincell{c}{$\begin{cases} \bm{d}^{E,z}_{1}=S_+(\bm{k})\bm{z}\\ \bm{d}^{E,z}_{2}=S_-(\bm{k})\bm{z}\end{cases} $  \\$\begin{cases}\bm{d}^{E,xy}_{1}=\sqrt{2}S(\bm{k})\bm{x}_+ \\
  			\bm{d}^{E,xy}_{2}=-\sqrt{2}S(\bm{k})\bm{x}_-\end{cases}$  \\$\begin{cases}\bm{d}^{E,\widetilde {xy}}_{1}=\sqrt{2}S_-(\bm{k})\bm{x}_-\\
  			\bm{d}^{E,\widetilde {xy}}_{2}=-\sqrt{2}S_+(\bm{k})\bm{x}_+\end{cases}$}\\
  		\specialrule{0.05em}{3pt}{3pt}
  		\hline
  	\end{tabular}
  	\label{tab1}
  \end{table}

    \begin{figure}[t]
    	\includegraphics[width=8.5cm]{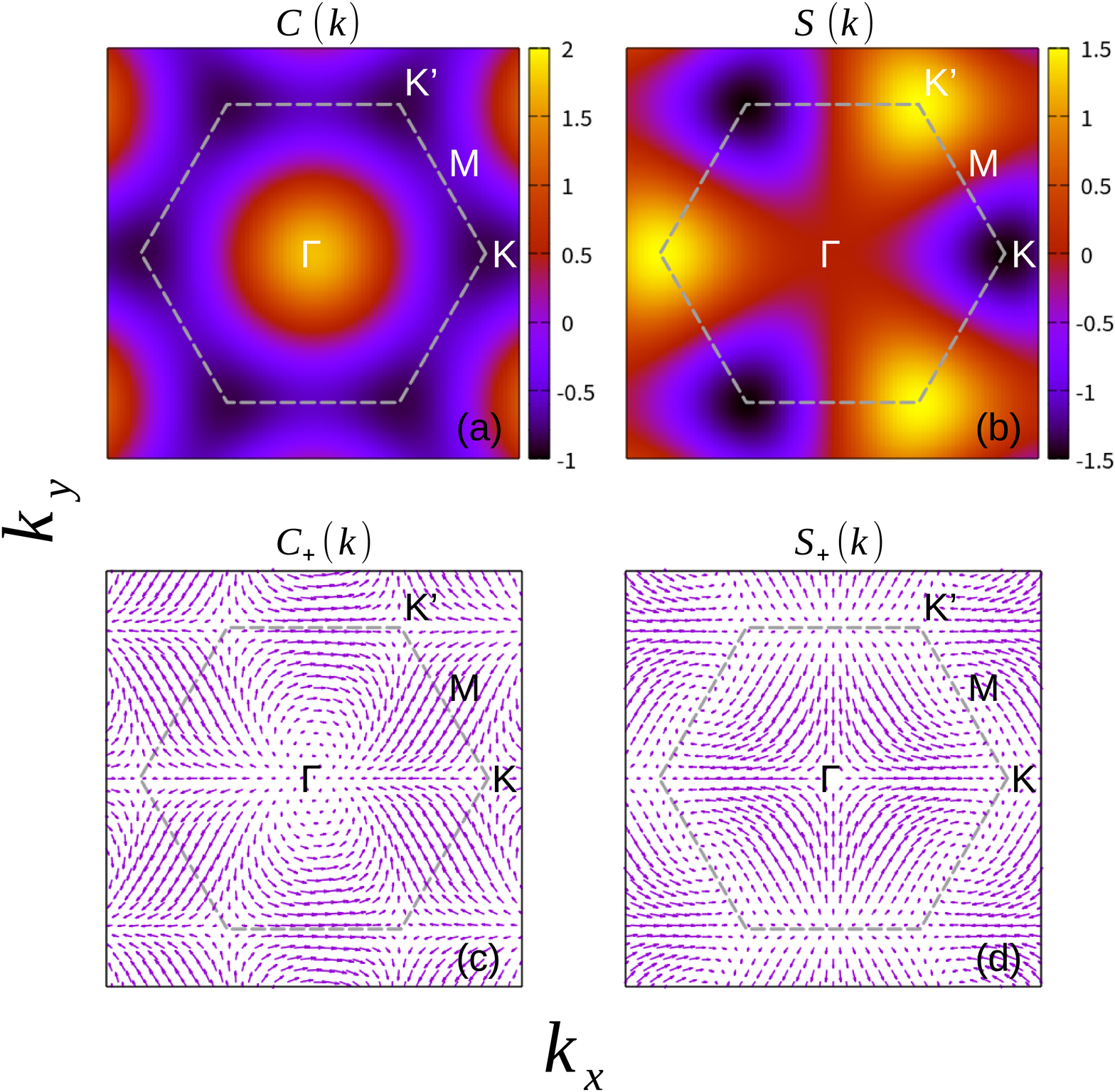}		
    	\vspace{-0.0cm}
    	\caption{\label{fig05}. Dispersions of the gap function (a)$C(\bm{k})$[extended s-wave], (b)$S(\bm{k})$[f-wave], (c)$C_+(\bm{k})$[d+id-wave] and (d)$S_+(\bm{k})$[p-ip-wave] in the momentum space. The hexagonal BZ is indicated by the gray dashed line. The orange and blue in (a)-(b) represent positive and negative values. The length and angle of arrows in (c)-(d) respectively give the amplitude and phase of $C_+(\bm{k})$, $S_+(\bm{k})$.} 
    \end{figure}

The Rashba spin splitting caused by the Mirror asymmetry of monolayer H-TaSSe can be written as
\begin{eqnarray}
&\bm{\alpha}_{\bm{k}}&=\alpha_{R} \{\frac{\sqrt{3}}{2}\bm{x}(\sin k_2-\sin k_3) \nonumber \\
&&-\bm{y}[\sin k_1-\frac{1}{2}(\sin k_2+\sin k_3)]\}
.\end{eqnarray}
The parameter $\alpha_R$ is taken from the H-WSSe\cite{Hu2018}, whose Rashba splitting strength is 158 meV$\cdot$\AA. This Rashba strength is corresponding to an energy 0.032 eV in our lattice model if we assume the lattice constant $a=3.3$ \AA, the same as in monolayer H-TaS$_2$\cite{Sanders2016}. The selection of $\alpha_R$  is reasonable for the atomic-structure similarity between tantalum and tungsten. 

These two kinds of SOC can be combined as $\bm{g_k}=\bm{\alpha}_{\bm{k}} +\bm{\beta_k}$, then the normal Hamiltonian of monolayer H-TaSSe is
\begin{eqnarray}
H_0(\bm{k})=\xi_{\bm{k}}+\bm{g_k}\cdot \bm{\sigma}
.\end{eqnarray}
  As a result, the spin splitting around $\Gamma$ is Rashba-type while around K it keeps Ising-type\cite{Hu2018}. The chemical potential $\mu$ of H-TaSSe is hard to be determined in our model, and its FS topology can be either similar to that of H-TaS$_2$ or H-TaSe$_2$. In view of this, both two cases are under our consideration and corresponding calculations are also made in section \ref{sec4}. The low-energy spectrum and FS of H-TaSSe at the two specific chemical potentials, $\mu=0$ and -0.18 eV, are simulated and shown in Fig. \ref{fig01}(b)-(c). It can be seen from the figures that the first Lifshitz transition occurs nearly at  $\mu=-0.15$ eV.  The ``Q" labels the position of the corresponding van Hove singularity(vHS) in the Brillouin zone(BZ) .

\section{Symmetry Analysis and Method}

The point-group symmetry of monolayer H-TaSSe is C$_{3v}$, basis gap functions of which are all presented in Table \ref{tab1}. Only the on-site and nearest-neighbor[NN] pairing channels are under our consideration. According to the spin angular momentum of the pair, i.e. 0[spin-singlet] or 1[spin-triplet], gap functions will take the matrix form as:

	\begin{eqnarray}\Delta^{\Gamma,\alpha}_{i}(\bm{k})=
	\begin{cases}
	\Psi^{\Gamma,\alpha}_{i} i\sigma_y & {singlet \ pairing}\\
	\bm{d}^{\Gamma,\alpha}_{i} \cdot \bm{\sigma} i \sigma_y & {triplet \ pairing}
	\end{cases}
	,\end{eqnarray}
where $\Psi^{\Gamma,\alpha}_{i}$ is the singlet order parameter and $\bm{d}^{\Gamma,\alpha}_{i}$ the $\bm{d}$-vector for the triplet pairing. Thus, the stable pairing of our system can be written as a linearized combination of these basis functions:
\begin{eqnarray}
\Delta(\bm{k})/\Delta=\sum_{\alpha}  c^{\Gamma,\alpha}\Delta^{\Gamma,\alpha}(\bm{k})
,\end{eqnarray}
where $\Delta $ is the gap size value and $c^{\Gamma,\alpha}$ denotes the relative amplitude of $\Delta^{\Gamma,\alpha}(\bm{k})$, with the orthogonal relation $\sum_{\Gamma,\alpha} |c^{\Gamma,\alpha}|^2=1$.

From Table \ref{tab1} we can get that $\Psi^{A_1,nn}_{}=C(\bm{k})$ represents the extended s-wave pairing, and $S(\bm{k})$ the f-wave pairing, while $C_{\pm}(\bm{k})$ and $S_{\pm}(\bm{k})$ denotes the chiral $d{\pm}id$ and $p_{\mp}ip$-wave pairings, respectively. Dispersions of these gap functions in the momentum space are shown in Fig. \ref{fig05}. SOC could destroy the inversion symmetry, giving rise to admixtures between singlet and triplet pairings. In addition, the $\bm{d}$-vector would tend to be paralleled to $\bm{g_k}$. So, in monolayer H-TaSSe with both strong in-plane Rashba and out-of-plane Ising SOC, the large admixtures between $\psi$ , $\bm{d}^{xy}$ and $\bm{d}^z$ are expected.

We capture the $T_c$ and the most stable pairing $\Delta(\bm{k})$ by solving the linearized gap equation:
\begin{eqnarray}
\Delta_{s_1s_2}(\bm{k})&=&T_c\sum_{\bm{k}^\prime s_3s_4}V_{s_1s_2s_3s_4}(\bm{k},\bm{k}^\prime)\nonumber \\ &&\times[G(\bm{k}^\prime,i\omega_n)\Delta(\bm{k}^\prime)G^{\tau}(-\bm{k}^\prime,i\omega_n)]_{s_3s_4}
,\end{eqnarray}
where $G(\bm{k}^\prime,i\omega_n)$ is the normal-state Matsubara Green's function, and $V(\bm{k},\bm{k}^\prime)$ the superconducting interaction. The form of $V(\bm{k},\bm{k}^\prime)$ can be given by the basis gap functions as follow:
\begin{eqnarray}
&&V_{s_1s_2s_3s_4}(\bm{k},\bm{k}^\prime) \nonumber \\
=&&-v_0\Psi^{A_{1},on}\Psi^{*A_{1},on}(i\sigma_y)_{s_1s_2}(i\sigma_y)_{s_3s_4}+  \nonumber \\
&&-v_1 \sum_{\Gamma,i}\Psi^{\Gamma,nn}_{i} \Psi^{*\Gamma,nn}_{i}(i\sigma_y)_{s_1s_2}(i\sigma_y)_{s_3s_4}+  \nonumber \\
&&-v_1 \sum_{\Gamma,\alpha,i} [\bm{d}^{\Gamma,\alpha}_{i}\cdot \bm{\sigma}i\sigma_y]_{s_1s_2}[\bm{d}^{\Gamma,\alpha}_{i}\cdot \bm{\sigma}i\sigma_y]^*_{s_3s_4}
,\end{eqnarray}
where $v_0, v_1$ represent the on-site and NN pairing constants and the positive(negative) values of them denote attractive(repulsive) interactions. Index $i$ is used to distinguish two equivalent basis pairing functions in the same IR. For 2D IR $\Gamma=E$, index $i=1,2$, while for 1D IR $\Gamma=A_1, A_2$, it is omitted.

\begin{figure*}[t]
	\includegraphics[width=17cm]{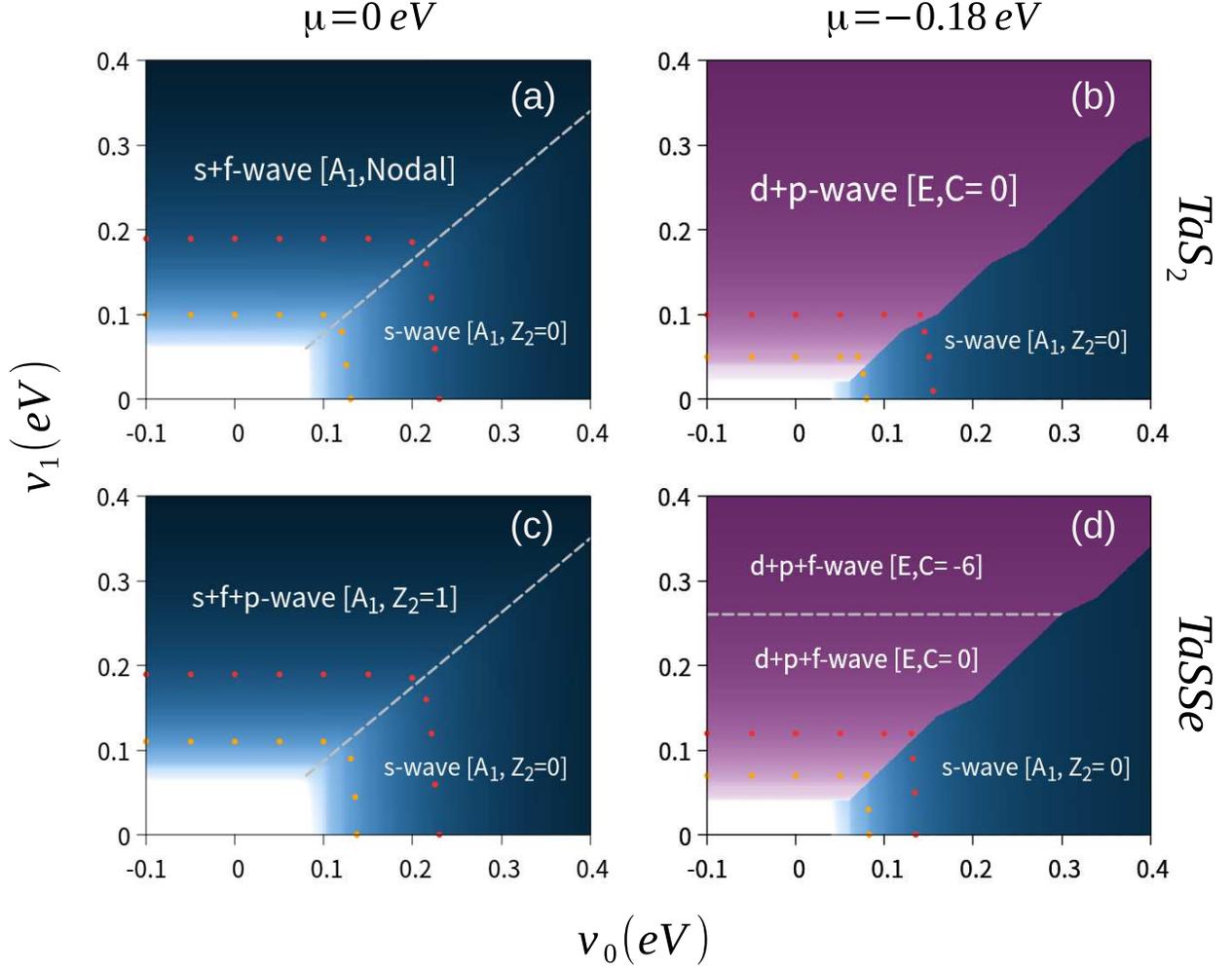}		
	\vspace{-0.0cm}
	\caption{\label{fig02}.
		The $v_0$ versus $v_1$ pairing phase diagrams of monolayer (a)(b) H-TaS$_2$  and (c)(d)of H-TaSSe at $\mu=0eV$ and $-0.18eV$. Phase $A_1$ is denoted by blue and $E$ by purple, while the dash indicates the borders between phases of the same IR. $T_c$ increases from bright to dark and the orange and red dots show equal T$_c$
		lines at 0.1 K and 10 K, respectively.
	} 
\end{figure*}

\section{Result and discussion}\label{sec4}

We obtain the $v_0$ versus $v_1$ pairing phase diagrams of H-TaSSe at specific chemical potential $\mu=0$eV and $-0.18$eV, which are presented in Fig. \ref{fig01}(c)(d). To investigate the strong Rashba effect on the pairing symmetry, the corresponding pairing phase diagrams of monolayer H-TaS$_2$ and H-TaSe$_2$[or can be viewed as a p-doped H-TaS$_2$] are also shown in Fig. \ref{fig01}(a)(b). 

As illustrated in Fig. \ref{fig01}(a), a nodal TRI mixed state $\Psi^{A_1,nn}+\bm{d}^{A_1,z}$ belonging to IR $A_1$, which is denoted as ``s+f-wave[$A_1$,Nodal]", dominates the phase diagram of H-TaS$_2$ if the NN pairing interaction is attractive and stronger than the on-site one. The s+f-wave pairing is a 2D Weyl superconductor and holds 12 nodal points on the $\Gamma$-centered FS\cite{Chen2019}. Otherwise, the conventional on-site s-wave has the largest gap amplitude. Border between the two phases is roughly depicted by the gray dash. 

In the H-TaSSe[see Fig. \ref{fig01}(c)], the large Rashba SOC induces helical p-wave component $\bm{d}^{A_1，xy}$ to the s+f-wave pairing, changing this nodal phase into full-gap. Typical relative amplitudes of these three components are $(c^{A_1,nn},c^{A_1,z},c^{A_1,xy})$=(-0.24,0.91,-0.34). The helical p-wave gap is larger than the extended s-wave gap on the $\Gamma$-centered FS. As a result, this mixed s+f+p-wave state is topologically nontrivial with a $\mathbb{Z}_2=1$ index\cite{Sato2009-1}.

\begin{figure}[t]
	\vspace{0.3cm}
	\includegraphics[width=8.5cm]{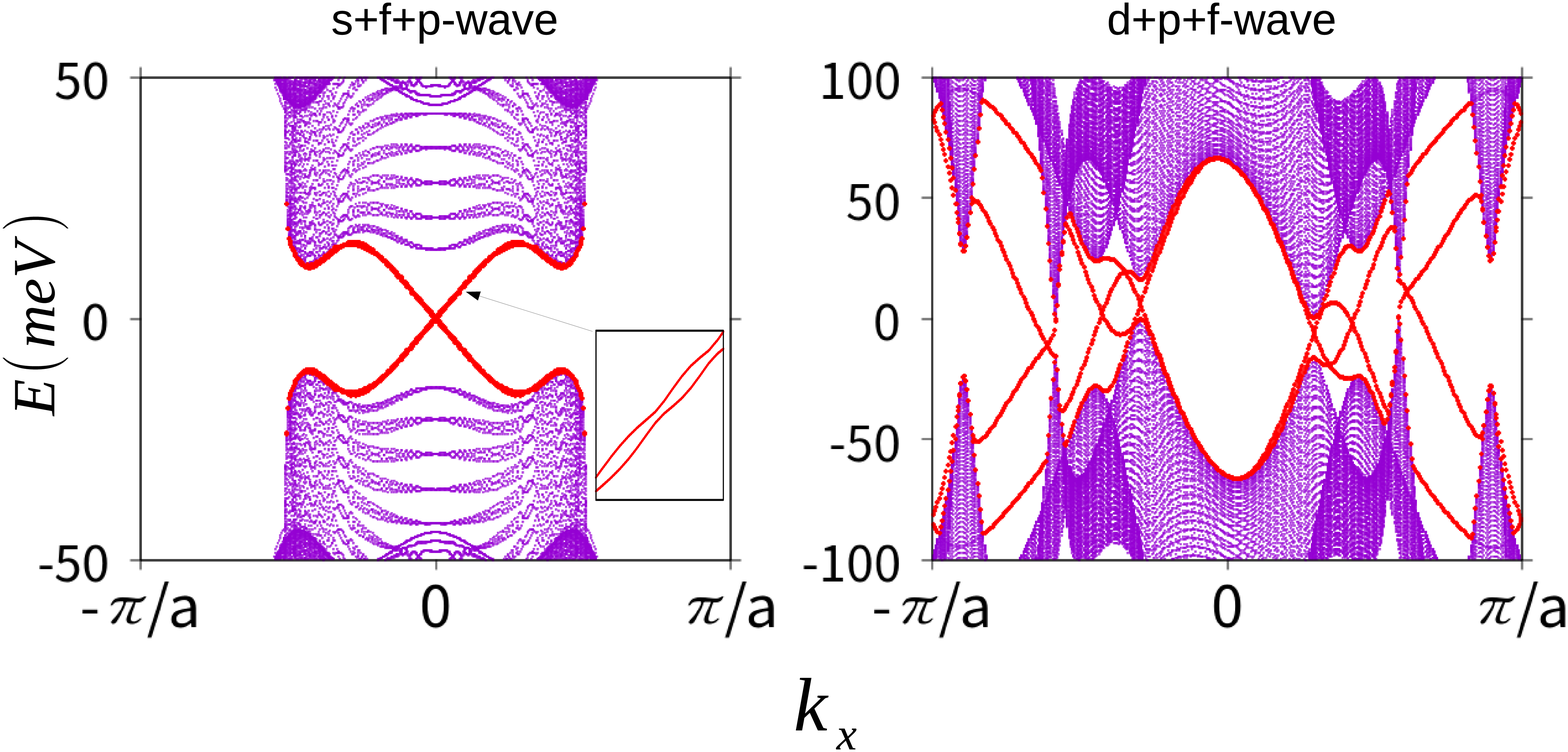}		
	\vspace{0cm}
	\caption{\label{fig03}.
		Energy spectra of the topological (a)TRI s+f+p-wave ($A_1, \mathbb{Z}_2=1$) state, (b)TRB d+p+f-wave ($E, C=-6$) state of the monolayer $H$-TaSSe, with open-boundary conditions along y direction. Gap size values $\Delta$ of them are enlarged to be about 10 meV for sight. The edge states are denoted by red.	} 
	\vspace{-0cm}
\end{figure}

In the p-doped H-TaS$_2$ with chemical potential $\mu=-0.18eV$, a mixed state $\Psi^{E,nn}+\bm{d}^{E,z}$[d+p-wave] of the 2D IR $E$ dominates the NN pairing channels[see Fig. \ref{fig03}(b)]. In principle, there are four components that actually can mix with each other: $\Psi_1^{E,nn}, \bm{d}_1^{E,z}, \Psi_2^{E,nn}$ and $\bm{d}_2^{E,z}$, whose relative amplitudes denoted by ($c_1^{E,nn},c_1^{E,z},c_2^{E,nn},c_2^{E,z}$). However, the minimization of the system's free energy gives the most stable d+p-wave mixed states with either $(c_1^{E,nn},c_1^{E,z})=0$ or $(c_2^{E,nn},c_2^{E,z})=0$. In another words, the admixture occurs only between d-id- and p+ip or d+id- and p-ip-wave. These two mixed states are equivalent and both break the time-reversal symmetry, allowing a calculation of Chern number\cite{Schnyder2008} on them. We adopt the d+p-wave pairing with specific amplitudes $(c_1^{E,nn},c_1^{E,z})=(-0.47, 0.89$), and the calculation gives a zero Chern number. In fact, the whole E phase in Fig. \ref{fig03}(b) is topologically trivial.

In the H-TaSSe with the same chemical potential, non-unitary pairings $\bm{d}^{E,xy}_{i}$ and $\bm{d}^{E,\widetilde {xy}}_{i}$ are induced to the $E$ phase. Based on the free-energy minimization, we choose $i=1$ for our following discussions. $\bm{d}^{E,xy}_{1}$ is an f-wave pairing emerging on the $\uparrow\uparrow$ pairing channel while  $\bm{d}^{E,\widetilde {xy}}_{1}$ a p-ip-wave pairing on the $\downarrow\downarrow$ channel. We denote this mixed state $\Psi^{E,nn}+\bm{d}^{E,z}+\bm{d}^{E,xy}+\bm{d}^{E,\widetilde{xy}}$ as the d+p+f-wave. It keeps topologically trivial in the diagram except the regions $v_1>0.27eV$, where  the nontrivial Chern number is $C=-6$, as shown in Fig. \ref{fig01}(d).  Typical relative amplitudes of the basis functions in the trivial and nontrivial cases are ($c_1^{E,nn},c_1^{E,z},c_1^{E,xy},c_1^{E,\widetilde{xy}}$)=(0.87,-0.39,-0.12,-0.28) and (0.8,-0.48,-0.14,-0.33), respectively. It is evident that a small modification of the ratios between these basis functions leads to a topological phase transition.

Comparing the phase diagram of H-TaS$_2$ to that of H-TaSSe, we can see that strong Rashba effect could change the topology of the NN pairing phase by inducing additional triplet pairings with vector $\bm{d} \perp \bm{z}$. Therefore, the possible topological pairing phase in H-TaSSe can be either an s+f+p-wave $\mathbb{Z}_2$ or a d+p+f-wave chiral mixed state depending on the chemical potential. On the edges, the former hosts robust helical Majorana zero modes(MZMs), while the latter has six channels of chiral ones. The corresponding energy spectrum of them are displayed in Fig. \ref{fig03}.
Recent superconducting conductance experiments observed zero bias conductance peak of the thin flakes of superconducting 2H-TaS$_2$ and 2H-TaSe$_2$\cite{Galvis2013,Galvis2014}, indicating novel superconductivity in the atomic limit of 2H-TMDs. Theoretical calculations suggest that the novel pairings of thin-layer 2H-TMDs could be a topological s+f-wave or d+p-wave\cite{Yuan2014,Hsu2017}, both of which require a NN pairing attraction larger than the on-site coupling. Consequently, assuming the same interaction relation $v_1>v_0$ here, it is very likely to realize the s+f+p-wave or d+p+f-wave pairing in monolayer H-TaSSe.

\begin{figure}[b]
	\includegraphics[width=8.5cm]{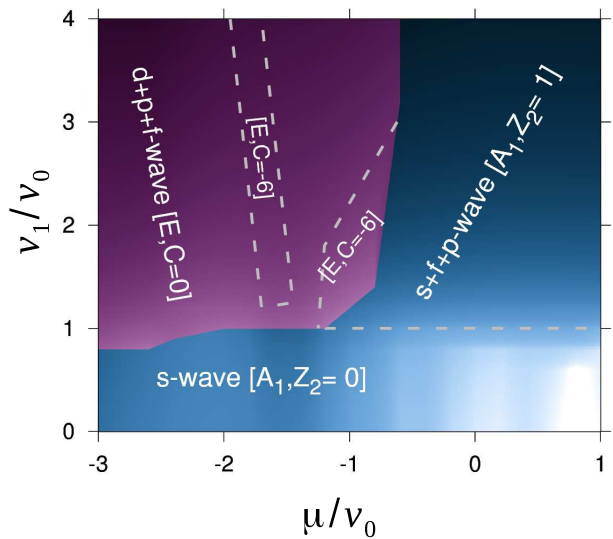}		
	\vspace{-0.2cm}
	\caption{\label{fig04}$v_1/v_0$ versus $\mu/v_0$ pairing phase diagram of monolayer  H-TaSSe, with $v_0=0.1$eV. The blue and purple denote $A_1$ and $E$ phase, respectively. $T_c$ increases from bright to dark.
	} 
	\vspace{-0.3cm}
\end{figure}

To further probe the pairings of monolayer H-TaSSe with respect to the chemical potential, we make a $v_1/v_0$ versus $\mu/v_0$ pairing phase diagram as shown in Fig. \ref{fig04}, with $v_0=0.1 eV$.  Consistent with the  above results,  once $v_0>v_1$, the most stable phase is the conventional s-wave. Otherwise,  it would be the NN pairings: The TRI s+f+p-wave pairing state dominates the parameter regions where $\mu>-0.08eV$, while the d+p+f-wave pairing is stabilized in other regions. As can be seen, when $\mu$ decreases, the $E$ phase emerges before the onset of the first Lifshitz transition[-0.15 eV]. This emergence of $E$ phase can be qualitatively understood as follows: At a high chemical potential, there are FS sheets centered at K(K'), where $S(\bm{k})$[f-wave], the dominant part of the s+f+p-wave, has the maximum gap value[see Fig. \ref{fig02}(b)]. As $\mu$ falls, the Fermi level will approach the saddle points M and Q, but away from K(K'). Gap functions $C+(\bm{k})$[d-wave] and $S+(\bm{k})$[p-wave], the main components of the $E$ phase, have gap-maximum at M and Q[see Fig.\ref{fig02}(c)-(d)]. Generally speaking, the larger gap on the FS, the higher $T_c$. Therefore, when $\mu$ decreases to about $-0.08$eV, the $T_c$ of $E$ and $A_1$ phases become equal and a phase transition emerges. Most of the $E$-phase regions in the diagram are topologically trivial, except two islands with Chern numbers -6. These two islands located nearly on both sides of the vHS[-0.15 eV], indicating that the FS topology of the normal H-TaSSe is not essential for the topology  of $E$ phase actually.

\section{Conclusion}
In this paper, the possible pairings of Janus monolayer H-TaSSe at different parameters have been investigated. By analyzing the pairing symmetry and calculating the linearized gap equation, we've identified a $\mathbb{Z}_2$ topological s+f+p-wave pairing state at the relatively high chemical potential, and a chiral d+p+f-wave pairing state with a large Chern number $C=-6$ at the lower one. The results show that Janus monolayer TaSSe can be a promising intrinsic helical or chiral TSC. Although here we merely focus on H-TaSSe, the conclusion can be applied to other superconducting Janus monolayer TMDs with a strong Rashba splitting. Our work will help to find the TSCs and realize the MZMs in the future.

\section{ACKNOWLEDGMENTS}
We thank X. Xun and H. Ya for useful discussions.
This work is supported by NSFC Project NO.111774126 and 973 Projects No.2015CB921202.

\bibliography{Janus}

\end{document}